Linear and Nonlinear Ultra-Short Pulse Looped Antennas: Radiation and Parametric Oscillations

H. Grebel
Electronic Imaging Center and the department of Electrical and Computer Engineering, NJIT, Newark, NJ, 07102. grebel@njit.edu

**Abstract:** Modern optical systems send and receive ultra-short temporal pulses (USP). While ultra-broad band antennas do exist in the microwave region (e.g., log-periodic antennas), their short temporal response is typically limited by the antenna's large dispersion, hence, resulting in a substantial pulse broadening. The issue becomes more severe when one considers both the transmitted and received pulses. Through simulations and experiments one can show that properly designed loop antennas, either thick loops or 3-loop antennas, exhibit USP attributes – 280 ps upon transmission and 380 ps upon reception (or, an overall equivalent coherent channel exceeding 2.5 GHz). Finally, most parametric amplifiers are narrow band and one may ask if a broadband amplification is possible. A loop inside a loop system, coupled by a nonlinear impedance element exhibits a line narrowing and signal amplification with large bandwidth, which is inversely scalable with the loops' diameter. In all, these elements could be advantageous for applications such as ultra-wide bandwidth communication and non-linear quantum information systems.

## I. Introduction

Ultra-short optical pulses (USP) are well-known for many applications; laser ablation, communication, light detection and ranging (LIDAR), in addition to many medical applications [1-2]. The RF/microwave versions is typically limited by large antenna dispersions. In the past, we showed that broad-side dipole array and two, side-by side diamond-shape antennas array may be constructed as to produce ultra-short pulses - as narrow as 100 ps (or, coherently, a broad bandwidth, larger than 10 GHz) [2-6]. Concentric loop antennas have been shown to possess a large bandwidth of ca 1 GHz [7-9]. Related applications with concentric antenna arrays (not to be confused with the shape of the antenna itself) have been proposed, as well [10-11]. Further applications of USP may be envisioned in ultra-wide bandwidth (UWB) systems [12] and in non-linear quantum information [13-15].

Parametric amplifiers are two oscillators coupled with a nonlinear element [16]. Commonly used couplers are nonlinear capacitors (varactors), or even a diode at the breaking or saturation points [17]. The transfer of energy from a strong pump at a frequency, 2f, to a weak signal at a frequency of, f, results in signal amplification. These amplifiers are employed in quantum information circuits, and specifically, in superconducting circuits, due to their low noise attributes [18]. As currently constructed, the amplification occurs at a relatively narrow frequency bands. We ask whether a broadband parametric amplifier is possible. The answer is yes if one considers line narrowing and amplification.

We concentrate on planar loop antennas which are amenable to conformations. The loop's width and the gaps between loops translate to the intra-coupling and affect the ability of the antenna to transmit and receive short pulses. We consider here a single (thick and thin) element and concentric 3-loop antenna to show that one can configure an antenna that can transmit pulses as short as, 260 ps and receive pulses as short as 380 ps.

Simulations for the various structures are provided in Section II: frequency domain (Section II.a); time domain for a monocycle and a temporal edge (Sections II.b); loops coupled by a nonlinear element (Section II.c). Experiments are provided in Section III: Frequency domain (Section III.a);

time domain (Section III.b) and the response to a monocycle (Section III.c). Discussions and Conclusions are provided in Section IV.

## II. Simulations

### II.a. Simulation at the frequency domain

All of our antennas are considered medium-size loop antennas, namely, the antenna's circumference is of the order of the transmitted/received wavelength. As a result, the antennas radiated perpendicularly to the loop plane. Three antenna structures are analyzed: the well-known thin, single loop, a solid, thick solid loop and a concentric 3-loop (Fig.1). The transmitter/receiver simulation cell is shown in Fig. 2. The simulations were conducted with a Comsol Multiphysics code. The transmitter/receiver pairs, shown in Fig. 2, are separated by 25 cm from each other to conform to the experimental setup. A single exciter of 1 V is placed horizontally at the transmitter bottom.

The various S-parameters for each antenna type are shown in in Fig. 3. A single-loop antenna (Fig. 1a) of diameter of 8.2 cm exhibits a relatively narrow-band radiation, of 150 MHz in the vicinity of $f_0$=1 GHz (Fig. 3a), $f_0$ being the center frequency. A thick solid loop (Fig. 1b) has the same outer diameter of 8.2 cm and an inner diameter of 5.2 cm). The antenna exhibits a much larger bandwidth of ~1 GHz between 1.1 to 2.1 GHz. A concentric 3-loop antenna (Fig. 1c) is basically a thick loop divided into three regions. It exhibits a relatively large bandwidth, as well: ~1 GHz between 1.25 to 2.25 GHz (Fig. 3c).

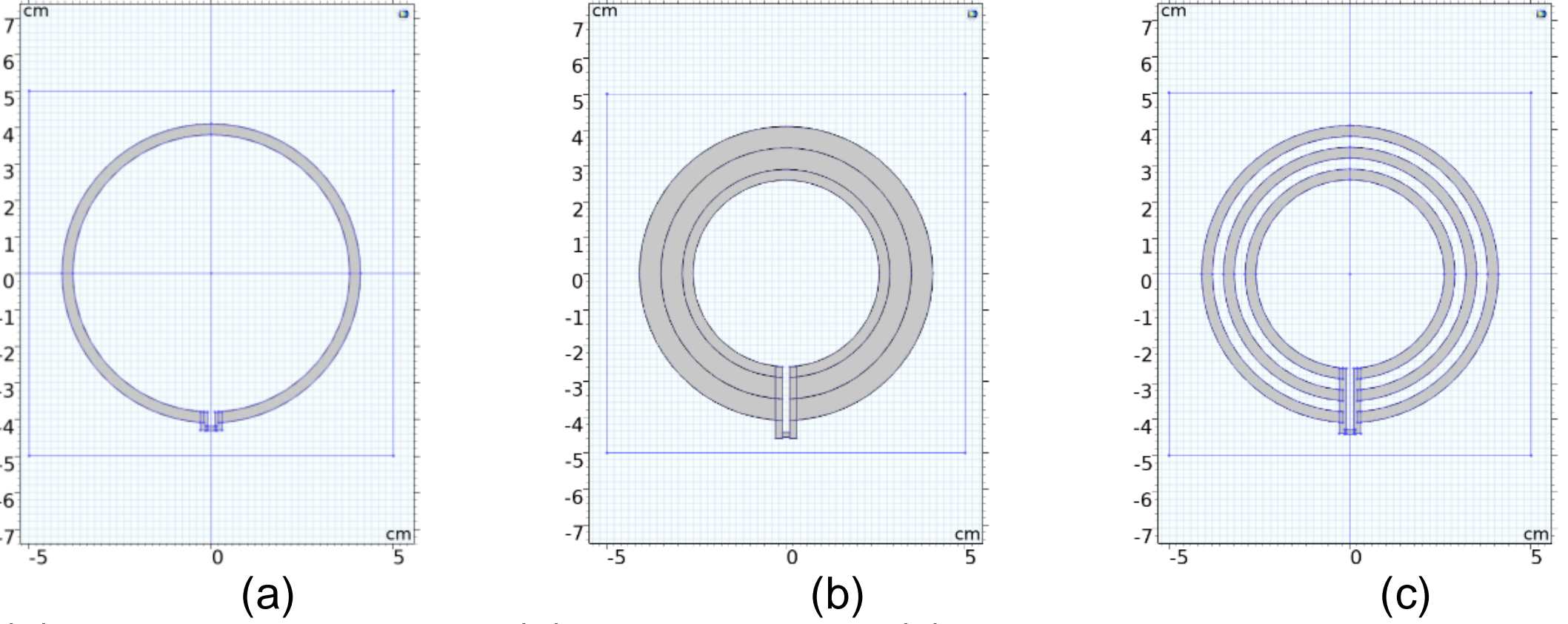

Figure 1. (a) A single loop antenna. (b) A thick loop and (c) a 3-loop antenna. All loops have the same outer diameter of 8.2 cm.

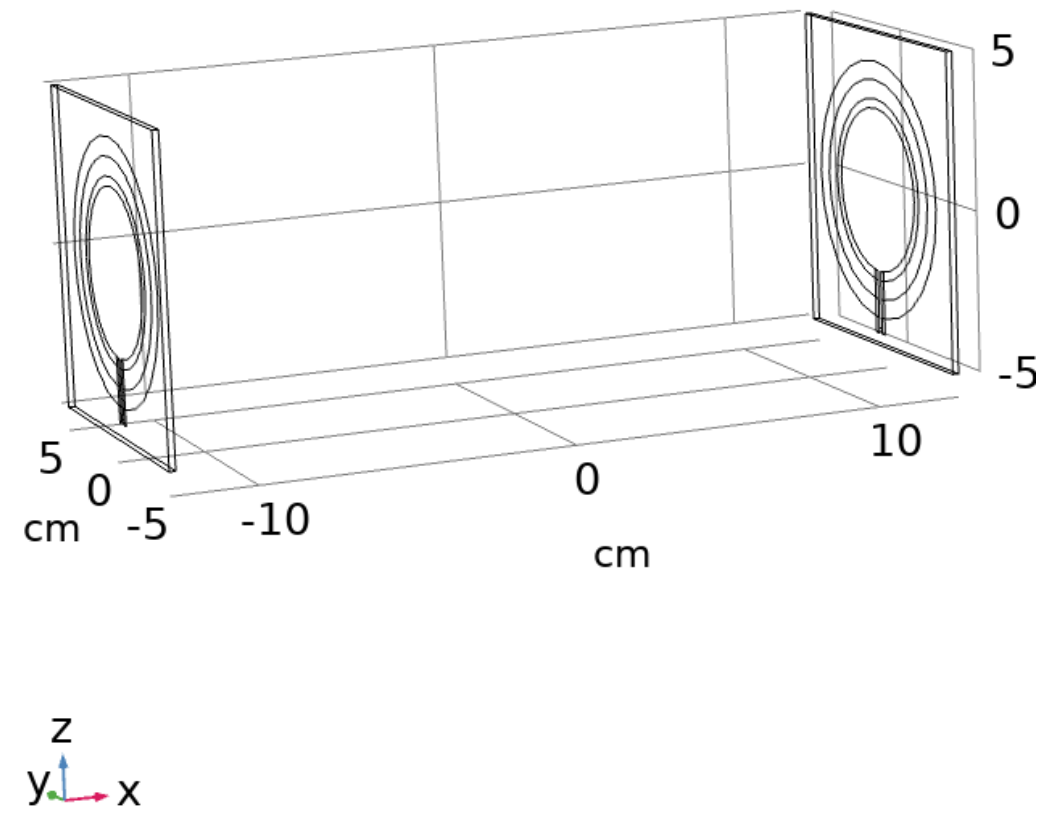

Figure. 2. Simulations cell. Shown is a transmitter and receiver pair situated at a distance of 25 cm from each other. The metallic antenna structure is placed on a standard printed circuit board.

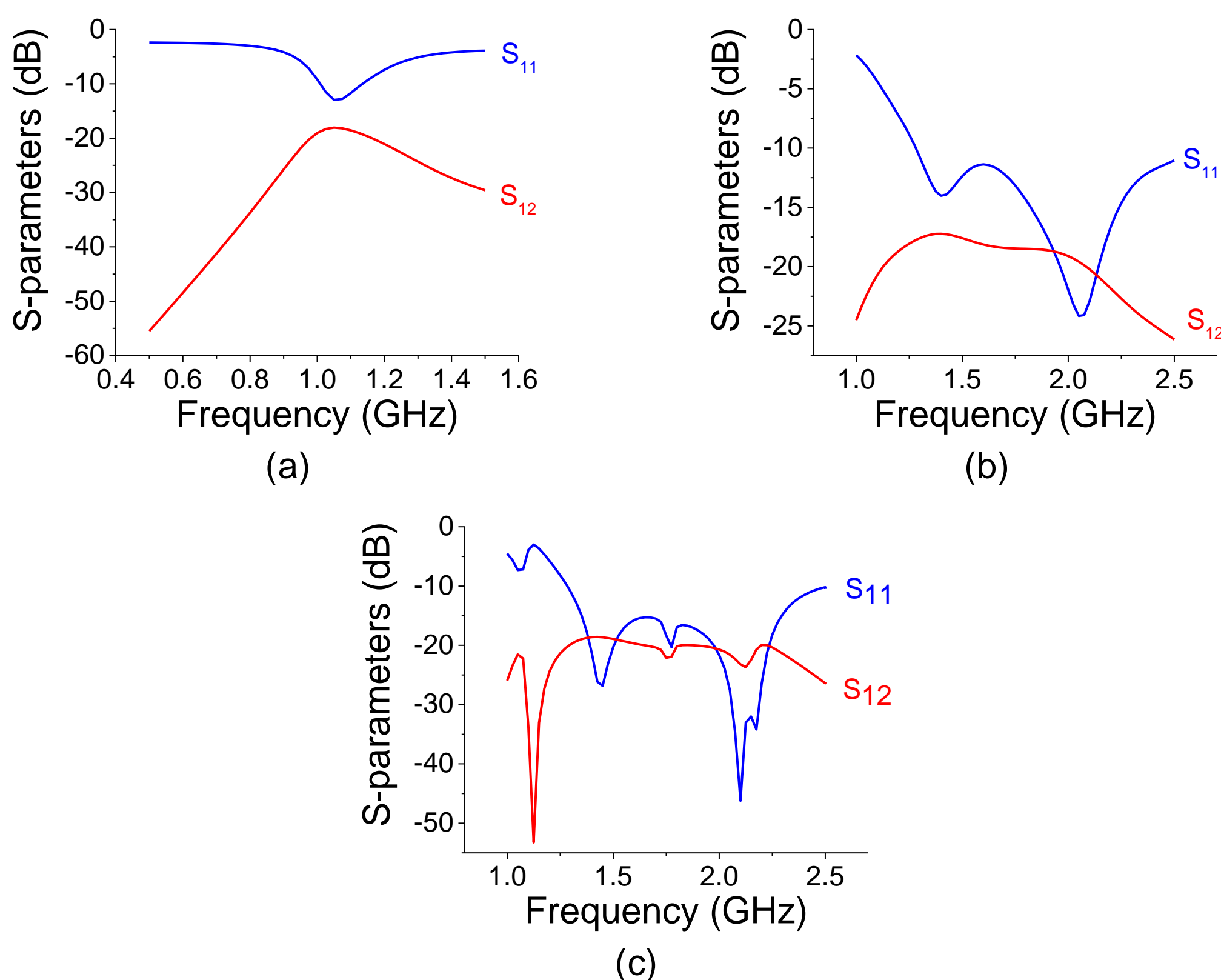

Figure 3. Simulations of S-parameters in dB for: (a) a single loop antenna; shown is the first order of scatterings; (b) thick loop and (c) 3-loops, equally spaced antennas.

As with other USP antennas, and unlike a broad-band frequency antennas, the gap between the conductors in a 3-loop antenna does not have a large impact on its temporal behavior. The small dips at f=1.7 and 2.1 GHz in the S-parameters of Fig. 3c shifts a bit with no effect on the SWR. In fact, a single thick loop (Fig. 3b) whose width is composed of the sum of all loops' widths (including the gaps) has a similar or better time response than the concentric 3-loop antenna as exhibited by the smoother $S_{12}$ parameter.

### II.b. Simulations at the time domain

### II.b.1. Monocycle pulses

As mentioned earlier, having a broad-band antenna does not guaranties formation of USP at the receiver end. The simulated temporal response to a transient - a Gaussian monocycle – had the form,

$$2 \cdot \left(\frac{A_s}{\tau}\right) \cdot \left(\sqrt{e}\right) \cdot (t - Tc) \cdot e^{-2[(t-Tc)/\tau]^2}). \qquad (1)$$

Here: t is the time; $A_s$ is the signal amplitude, here, $A_s$=1; $\tau$ is the cycle width, in our case, $\tau$=0.15 ns; Tc is the delay, in our case, Tc=0.5 ns. These parameters are similar to the 4 GHz monocycle experiments (see below).

The simulation results are shown in Fig. 4. Fig. 4a exhibits the response of a thick loop antenna. The input width (at FWHM) is 0.36 ns whereas the output from the receiver is 1.5 ns. The receiver output has a ~3 ns delay from the transmitter due to the 25 cm distance between them. The receiver shows ringing which could be due to higher-order round trip, since the cycle repeats itself every odd number of delay. This ringing is absent from the experimental data. The 3-loop antenna seems to suppress the first response at ~3 ns while both the transmitter and receiver signal have been stretched.

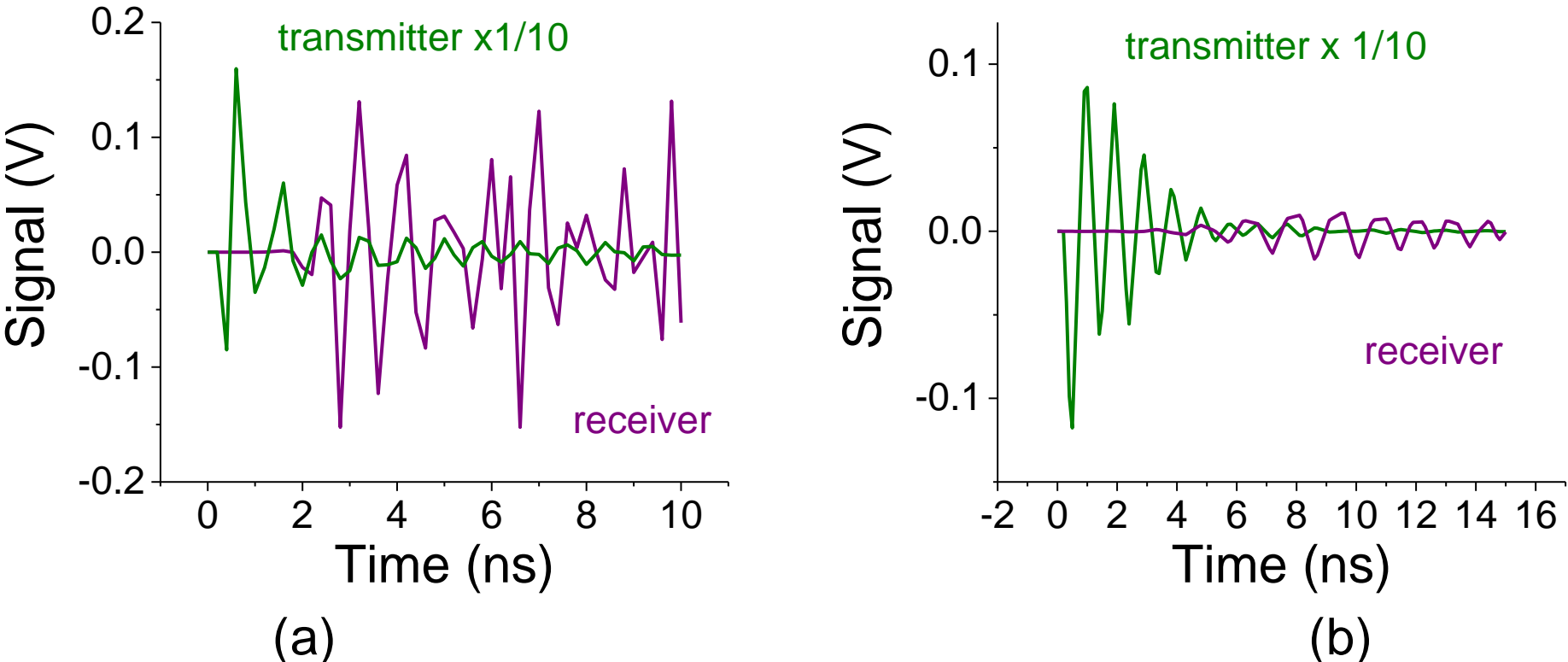

Figure 4. Simulations. Response to a Gaussian monocycle. (a) Thick loop antenna. (b) 3-loop antenna.

### II.b.2. Response to a temporal edge

Response to a sharp temporal edge is shown in Figs. 5a-c. The single-loop antenna cannot sustain a sharp input transient into the transmitter. This is also exhibited by the receiver response. On the other hand, the concentric 3-loop antenna exhibits both good input response by the transmitter and a reasonably good output response by the receiver. The input current in port 1 (the transmitter) is generated by a 1 V bias over 50 Ω input impedance of a transmission line that is connected to the transmitter; the output current in port 2 (the receiver) is measured at the end of the 50 Ω transmission line of the receiver. The distance here, between the input signal (transmitter) and the output signal (receiver) was a bit shorter than 25 cm – it was 16 cm, hence the delay between the input and output signal is a bit shorter; ~2.5 ns for the latter vs ~3 ns for the former. Again, the ringing follows roughly odd number of delays between the antennas and is absent from a single loop and also is absent from the experimental results.

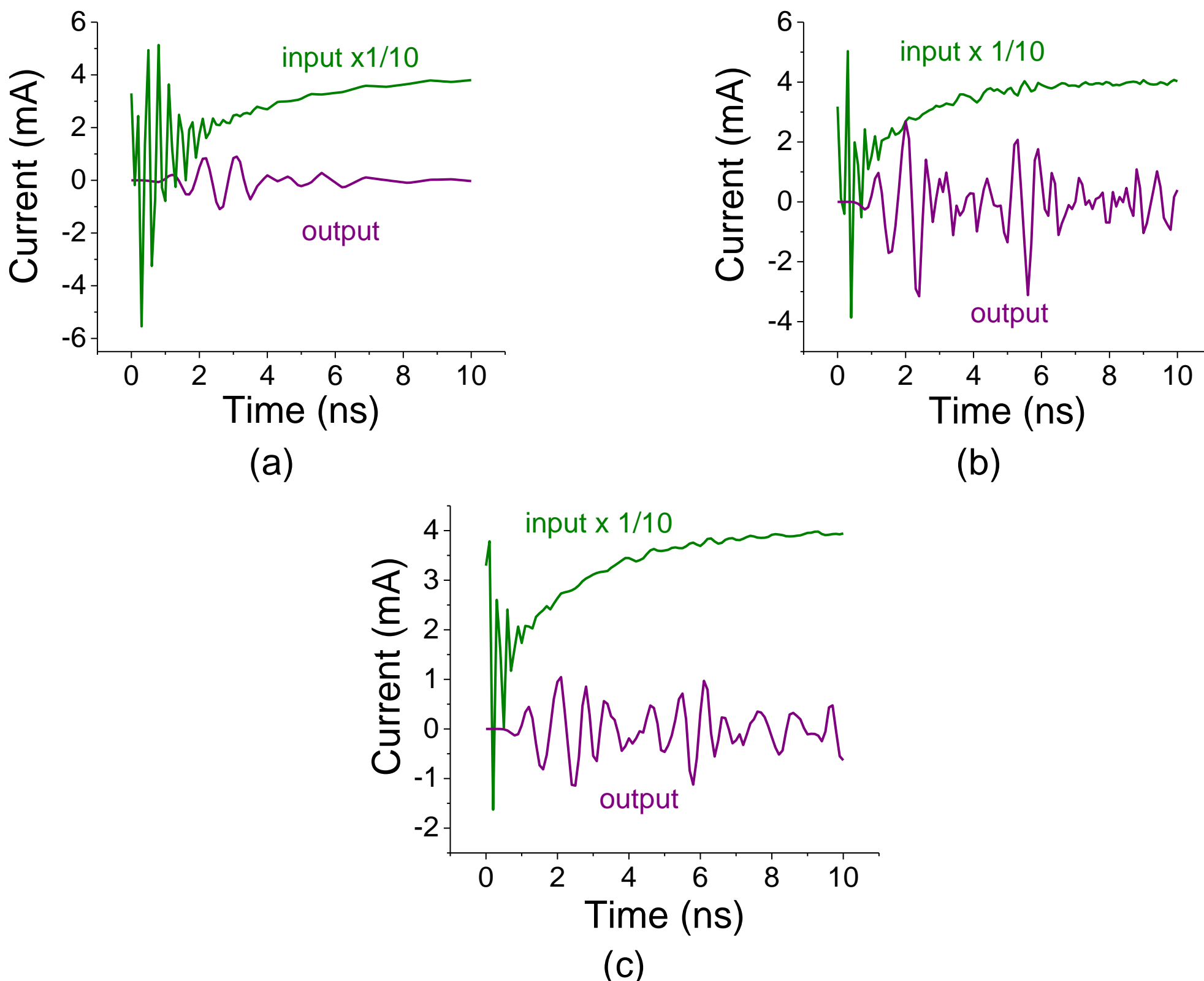

Figure 5. Temporal edge simulations: (a) a single thin loop antenna – input current (at the transmitter) and output current (at the receiver). (b) A thick loop antenna and (c) 3-loop antenna – input current (at the transmitter) and output current (at the receiver).

**II.c. Simulations with nonlinear coupling element**

Let us consider two thick loops. The outer diameter is twice the diameter of the inner loop as shown in Fig. 6.a., of diameters of 8.2 cm and 4.1 cm, respectively. The input monocycle signal is provided to the inner loop through a 50 Ohms transmission line. The output signal is taken from the outer loop through another 50 Ohms transmission line. The nonlinear element that couples the two loops is made of a homogeneous strip whose impedance is changing as:

$$A_p \cdot \mathrm{e}^{-[(t-Tc/2)/\tau_1]^2}. \qquad (2)$$

Such impedance change may be provided by a nonlinear resistor, memristor, or a superconducting element driven across it phase transition by say, an electromagnetic pulse at ca 40 GHz. As it turns out, the amplitude, $A_p$, has to be small and we allow the Gaussian pulse to start earlier than, and wider than the monocycle of Eq. 1. Fig. 6c,b exhibits the current drawn from the output port (the port attached to the larger diameter loop). Clearly, the nonlinear coupler provides for line narrowing (almost to a delta function), which alludes to wide band of the output signal. It also alludes to the amplification of the output signal. The delay in the signal is attributed to the signal build-up in the coupled resonating structures. The parameters used for both equations 1 and 2 were: Tc=0.5 ns; $\tau$=0.15 ns and $\tau_1$=0.3 ns.

In order to complete the picture we add power distribution in Fig. 6d.e. The delay in the coupled signals and their amplification is shown in the power distribution, as well. Assessment of the received pulse delay is provided in the Discussion Section.

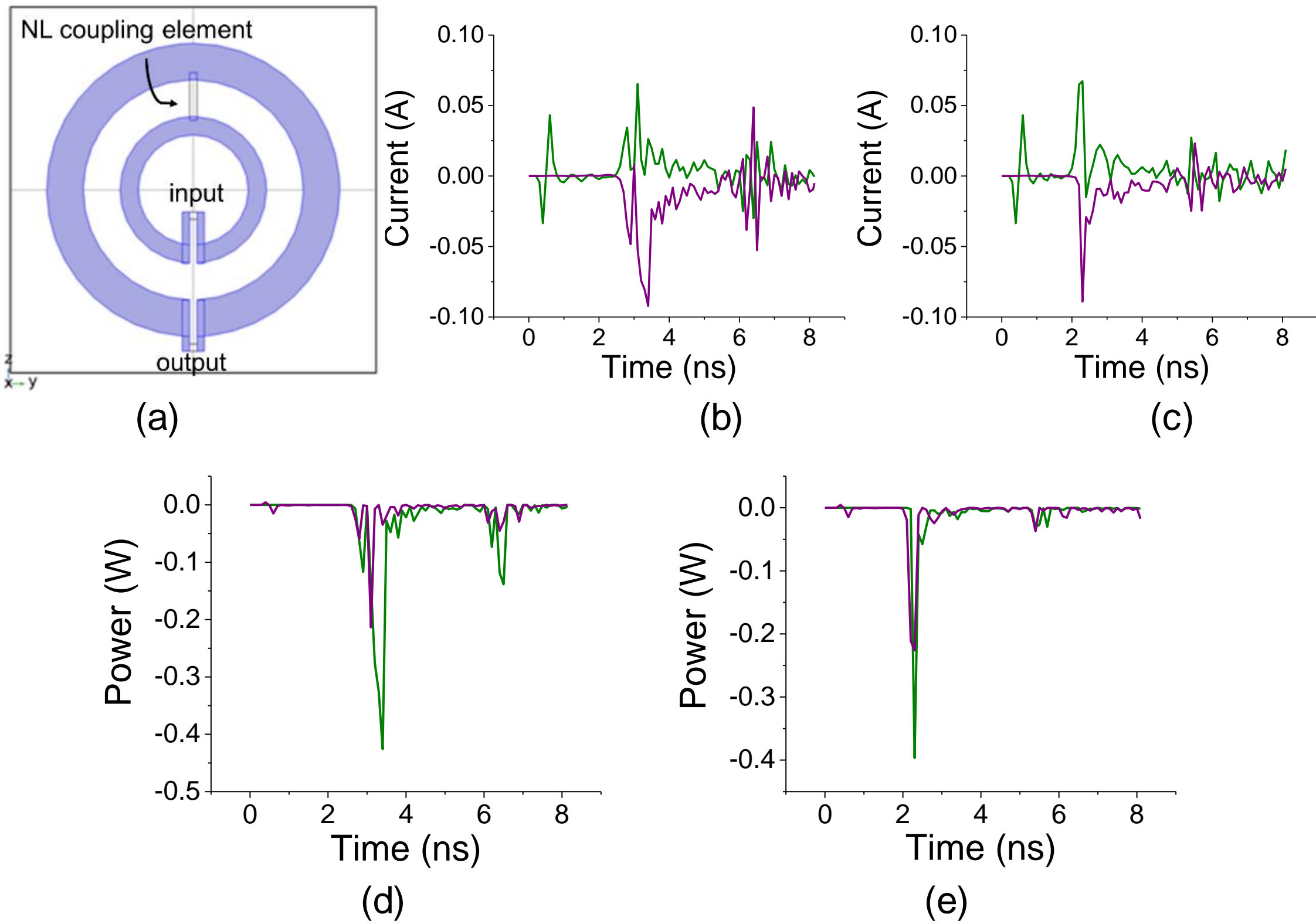

Figure 6. (a) Two thick loops (blue color), one within each other, of respective diameters, 8.2 cm and 4.1 cm. (b) Pump with A=10: input (green) and output (purple). (c) Pump with A=0.1: input (green) and output (purple). For (b) and (c), the original monocycle appears at around time, t=0.5 ns. (d,e) Power distribution for A=10 and A=0.1, respectively: input (green) and output (purple).

### III. Experiments

A pair of 3-loop antennas have been fabricated by a lift-off process resulting in Fig. 7a. The antenna pair were placed, facing each other at a distance of ca 25 cm. A 1 MHz to 8 GHz VNA (Copper-Mountain Technologies) was used. The system has a time domain capabilities. For the monocycle experiments, we used a 4 GHz monocycle unit (Avtech). Here, the VNA was triggered by the monocycle unit and the readings were taken either as a direct signal to the VNA (from port 2), or, from the VNA connect, receiver antenna (connected to port 1). For the thick loop antenna, the 3-loop were covered with a copper tap as shown in Fig. 7c.

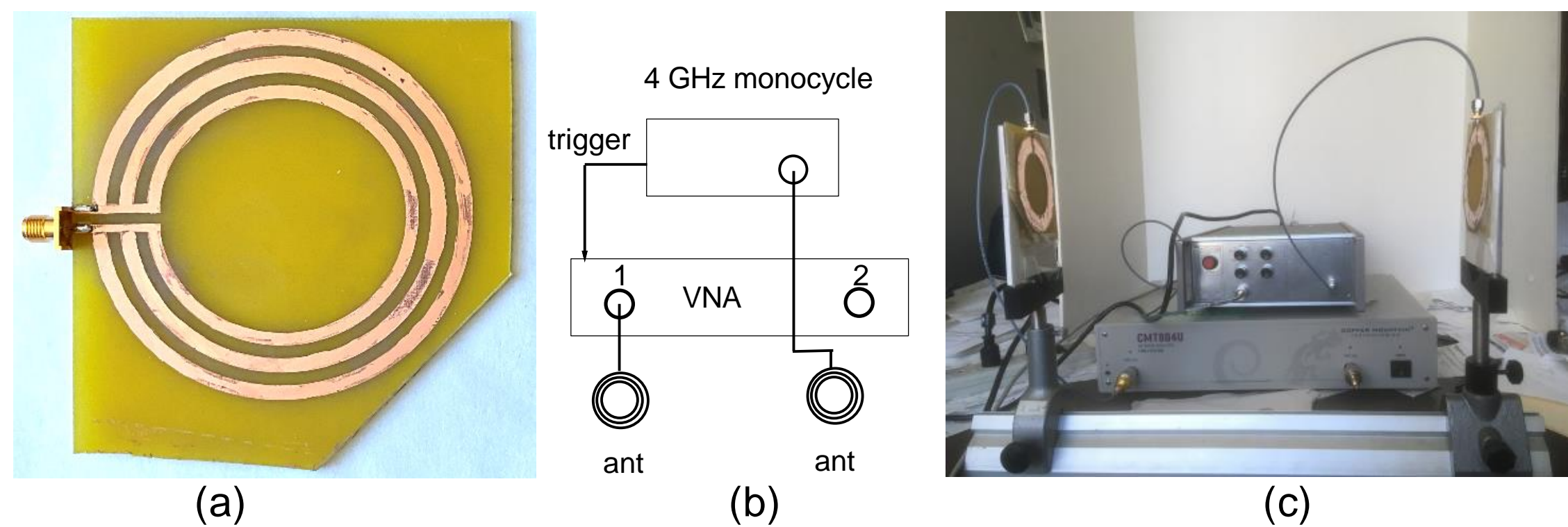

Figure 7. Experiments: (a) 3-loop antenna on FR4 dielectric. The cut in the FR4 is the result of fabricating two antennas on the same substrate. (b) Schematics of measuring a monocycle with VNA. (c) Two thick loop antennas facing each other for frequency and time domain experiments. The as-is reading of the monocycle is obtained by connecting the unit to Port 2. For the transmission measurements, the monocycle unit is connected to the transmitter antenna and the signal is detected by the receiver through port 1.

### III.a Frequency-domain results:

The S-parameters for a 3-loop antennas are presented in Fig. 8a. The experimental standing wave ratio (SWR) is shown in Fig. 8b. Both results corroborate the simulations.

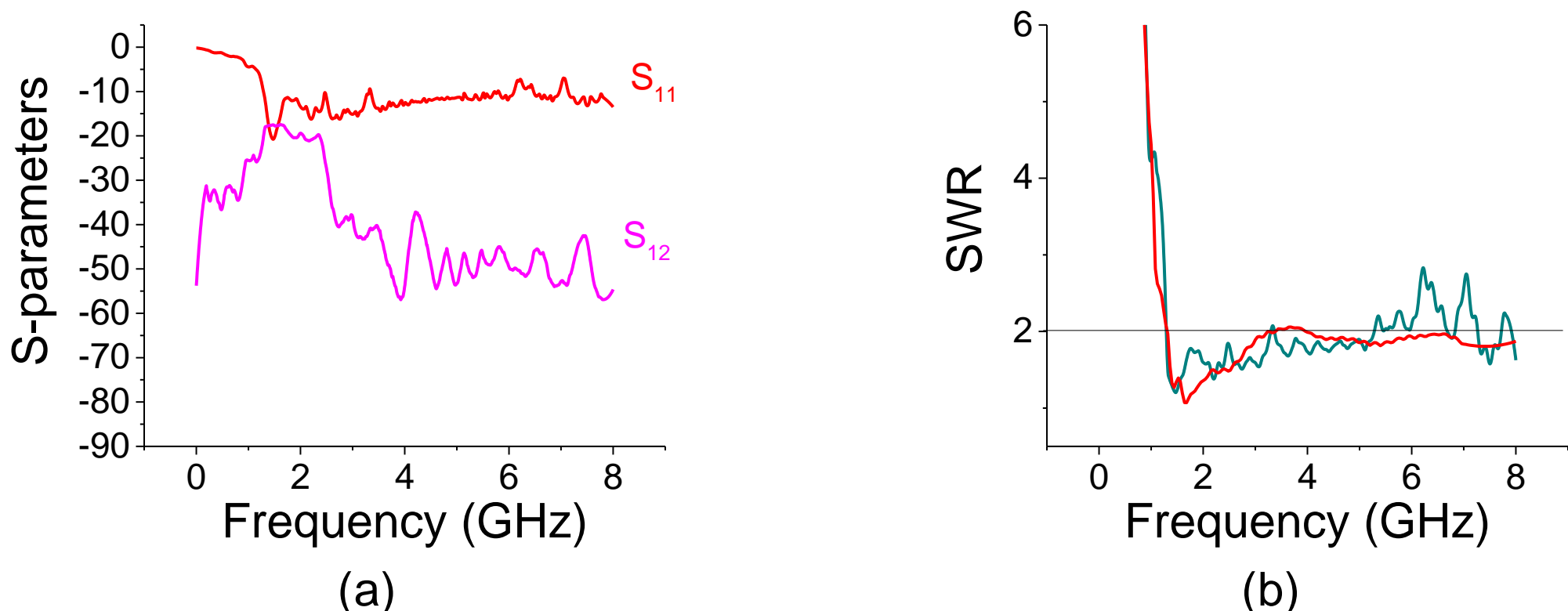

Figure 8. Experiments 3-loop antenna: (a) (smoothed) S-parameters in dB and (b) (smoothed) Standing wave ratio (SWR) of $S_{11}$. The red curve is for a thick loop antennas

### III.b Time-domain results:

Experimental time domain results are presented in Fig. 9. Shown are the linear magnitude of the S-parameters. There is a relative delay between pulses of ca 1 ns due the distance between antennas. The linear magnitude of $S_{11}$ exhibits a full width at half maximum amplitude (FWHM) of 260 ps for the total signal that includes the entire pulse. The linear magnitude of $S_{12}$ exhibits a FWHM of 380 ps.

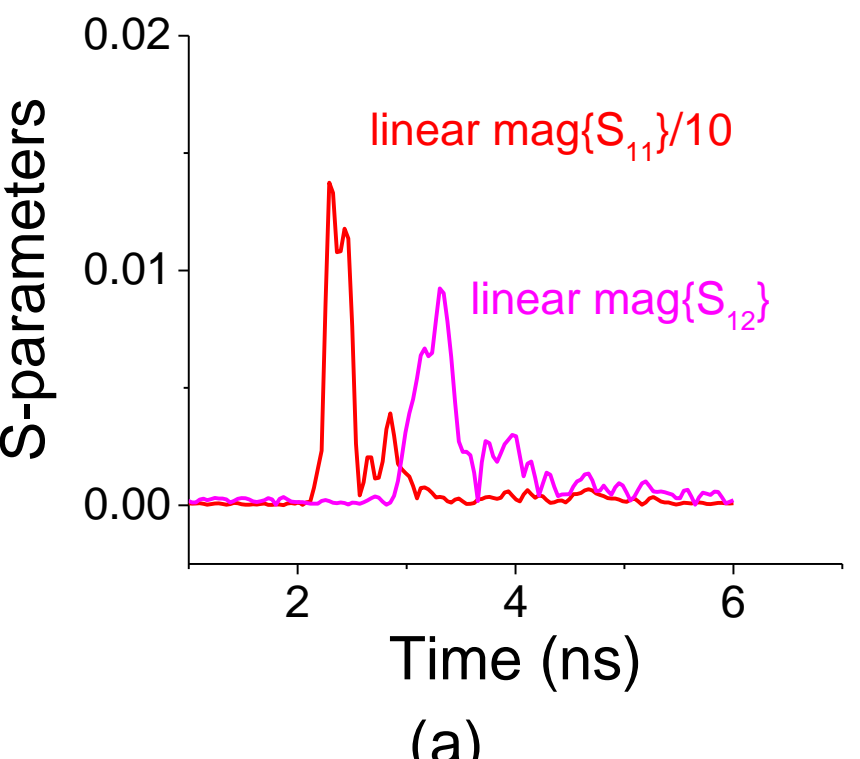

(a)

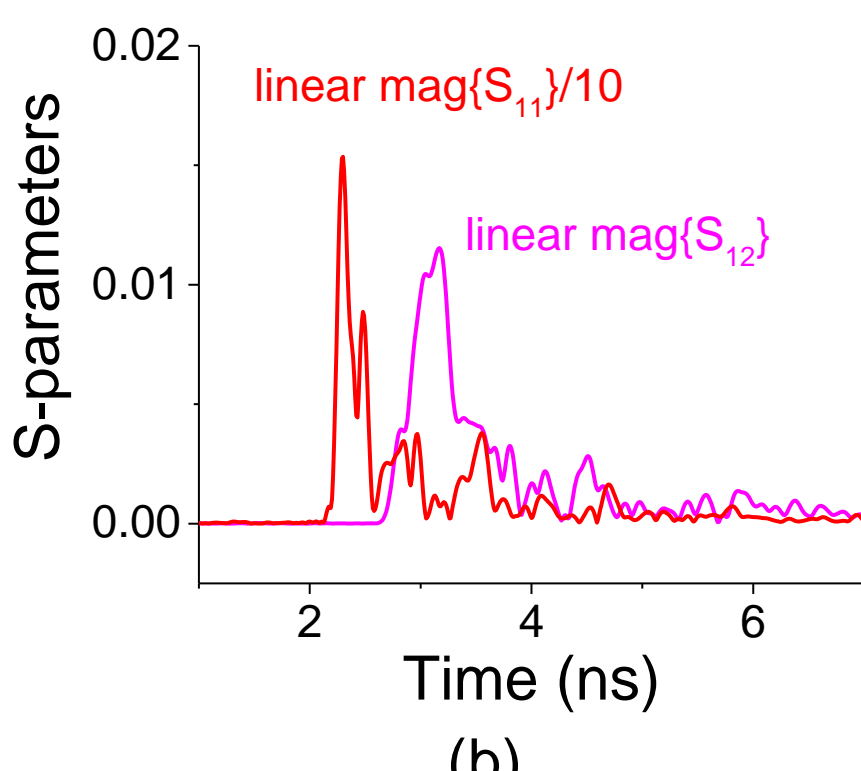

(b)

Figure 9. Experiments exhibiting linear magnitude of the S-parameters as a function of time (one way). (a) Thick loop: full width of the signal at half maximum amplitude (FWHM) of the $S_{11}$ was 260 ps and the $S_{12}$ was 380 ps. (b) 3-loop antenna: FWHM of the $S_{11}$ was 260 ps and the FWHM of the $S_{12}$ was 380 ps, as well.

The broadening of the received signal is related to a non-planar wave that propagates between the two close-proximity antennas in addition to the effect of the transmission line(s) (TL). Nevertheless, even with this pulse broadening, the coherent bandwidth of the entire transmitter/receiver system is ca 2.6 GHz.

### III.c Time-domain results: monocycle

Transmitted and received monocycles are shown in Fig. 10. The FWHM fed to the transmitter is 0.17 ns whereas the FWHM of the received signal is 0.6 ns. A delay of 5 ns in the received signal is also noted: it includes free-space delay (0.8 ns) and cable-delay of 3.8 ns (between the receiver antenna and the VNA port). Note also that the received signal is interpreted as $S_{11}$ by the VNA.

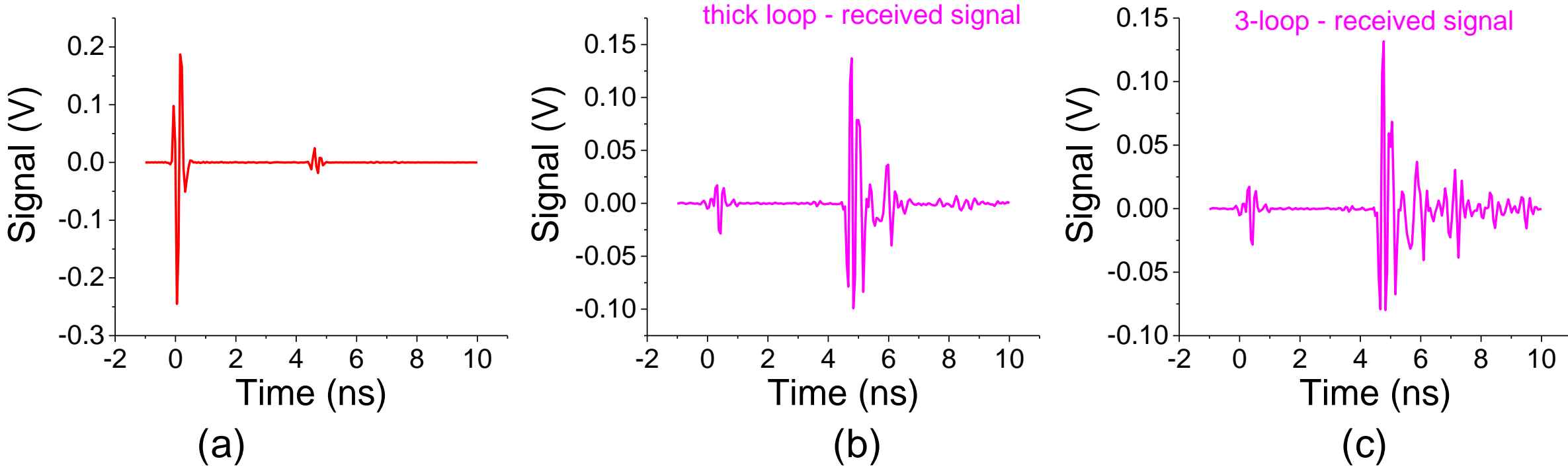

(a) (b) (c)

Figure 10. Experiments – monocycles: (a) input signal from the generator directly to the VNA. The 4 V amplitude is attenuated by 20 dB; (b) Received signal by a thick loop; (c) received signal by a 3-loop. The receiver was placed at 25 cm from the transmitted antenna.

## IV. Discussions and Conclusions

Loop antennas are planar structures and their characteristics in the time domain were analyzed and measured. Thin loop antenna is known to exhibit a narrow bandwidth of ca 150 MHz, as we found out and has served here as a reference.

Simulations – frequency and time domains: Thick-loop antenna, where the thickness-to-outer diameter ratio is larger than 1/3=0.333 – in our case the ratio was 3/8.2=0.365 – exhibited a smoother and broad $S_{12}$ curve and relatively larger transmission amplitude than its 3-loop counterpart. Time-response to a temporal edge exhibited a better performances with the thick loop antenna over the 3-loop in both transmission and reception.

Experiments – frequency and time domains: unlike the simulations, experimentally, both antennas exhibited similar time-domain response: the linear magnitude of the S-parameters as a function of time (one way) a full width of the signal at half maximum amplitude (FWHM) of the $S_{11}$ and $S_{12}$ as 260 ps and 380 ps, respectively. One should note that these time responses are based on Inverse Fast Fourier Transform (IFFT) that is performed by the VNA on the frequency domain data; this attests to a coherent frequency response (phase and amplitude) – otherwise, the IFFT would produce a broadened signal, which is not the case here.

Monocycles - Simulations: simulation of the antennas' response to Gaussian monocycle exhibited an input signal width (at FWHM) of 0.36 ns whereas the receiver signal has exhibited a widths of 1.5 ns. The receiver output has a ~3 ns delay from the transmitter due to the 25 cm distance between them. The simulations appeared to have shorter duration response for the thicker antenna. The ringing in the received signal was shorter and their amplitude stronger for the thick-loop antenna than the 3-loop antenna (Fig. 4).

Monocycles – Experiments: experimentally, the ringing effect has disappeared probably to channel losses. The FWHM fed to the transmitter was 170 ps whereas the FWHM of the received signal was 600 ps. While the received signal to the 3-loop antenna was nosier compared with the thick-loop antenna, the overall FWHM and their respective amplitudes were similar.

Nonlinear simulations: we provided simulations for two loop antennas coupled with a nonlinear low impedance element. The element can be realized by a nonlinear resistor (memristor); or, its nonlinearity may be provided by a transition between a superconducting/normal states (for example by a high-frequency electromagnetic pulse). A substantial line narrowing and enhancement of the output signal were exhibited. The delay in the received signal was a function of the NL element's impedance; it was smaller for the lower impedance coupler (Fig. 6c. vs Fig 6b.). The delay of the received signal is attributed to the onset of resonance conditions in both loops (see below). This mechanism may serve as a delay and power control by the pump pulse amplitude: as the impedance gets smaller, the fluence (energy/unit time) increases.

In Fig. 6.d,e we see that the power follows the current trend as far as the relations between a smaller delays, pulse narrowing and the received pulse amplification. The energy by the pump (which eventually is transferred to the weak signal) is indirectly implied by the impedance characteristics of the non-linear coupler. By Fourier transforming the received signal, one may conclude that shorter delays correspond to a broader frequency parametric bandwidth; for the specific case here with its 2 ns delay, we assess the parametric bandwidth as 160 MHz.

Pulse delay has to do with the onset of resonance in the two loops. The effective dielectric constant of the surface metal loop is $\varepsilon_{metal}=(\varepsilon_1+\varepsilon_2/2)$ where $e_{1,2}$ are the dielectric constants of the substrate and air, respectively. Here, $\varepsilon_1=4$ and $\varepsilon_2=1$ and $\varepsilon_{metal}=2.5$. The index of refraction of the metal is $n_{metal}=\sqrt{2.5}$. Thus, the phase velocity (ignoring dispersions) is $v=3x10^8/n_{metal}$. The onset of a standing wave in both resonators is the sum of their circumference length and the NL element length, of ca 1 cm. Thus, since the diameter of the resonators are, ca 8 and 4 cm, respectively, $L\sim\pi\cdot(8+4)+1$ cm. The delay time is $L/v\sim(38.7/1.9)\cdot10^{-10}\sim2$ ns, in reasonable agreement with the simulations. The results inversely scale with the loops' radii - the smaller the loops are, the higher are their resonating frequencies and the larger are their parametric bandwidth.

All in all, linear coupled loops antennas exhibit short pulse characteristics and its NL version exhibits signal amplification. These elements could be advantageous for applications such as ultra-wide bandwidth communication and non-linear quantum information systems.